\documentclass[11pt,eqsecnum]{article}

\usepackage{amsmath, amssymb}
 \usepackage{epsfig}
\usepackage{mathrsfs}
\usepackage{psfrag}
\usepackage{graphicx}
\usepackage{epstopdf}

\usepackage{color}

\usepackage{color}

\newcommand \be{\begin{eqnarray}}
\newcommand \ee{\end{eqnarray}}

\newcommand{\ba}{\begin{eqnarray}}
\newcommand{\ea}{\end{eqnarray}}

\def\l {\lambda}

\def \td {\tilde}
\def \td {\tilde}
\def\sn{{\rm sn}}

\def \g{\gamma}

\def\S{{\cal S}}

\def\k{\kappa}
\def\E{{\cal E}}

\def\s{\sigma}

 \def\tg{\tilde{\g}}

 \def\J{{\cal J}}
\def \adss  {$AdS_5 \times S^5$}
\def \sql {\sqrt{\lambda}}

\begin{document}
\renewcommand{\thefootnote}{\arabic{footnote}}

\def \foot {\footnote}
\def \bi{\bibitem}

\def \tr {{\rm tr}}
\def \ha {\frac{1}{2}}

\def \ci{\cite}
\def \N {{\mathcal N}}
\def \const {{\rm const}}
\def \t {\tau}
\def\S{{\mathcal S} }
\def \nn{\nonumber\\}
\def \XX {{\rm X}}

 \def \vp {\varphi} \def \bs {\bar \s }
\def \k {\kappa}
\def\foot{\footnote}
\def \ci {\cite}

\def \bp {\begin{pmatrix}}  \def \epm {\end{pmatrix}}
\def \ha {{\textstyle{1 \ov 2}}}

\def \bi {\bibitem}
\newcommand{\lar}{\longrightarrow}
\def \la {\label}
\def \Tr  {{\rm Tr}}

\def \T {{\cal T}}
\def \l {\lambda}
\def\foot{\footnote}
\def \tl  {{\tilde \l}}
\def \sql {{\sqrt \l}}
\def \adss {$AdS_5 \times S^5$\ }
\newcommand{\rf}[1]{(\ref{#1})}

\def \bp {\begin{pmatrix}}
 \def \emp {\end{pmatrix}}

 \def \dett  {{\det}}

\def \qr {{\hat \rho}}
\def \const {{\rm const}}
\def \bea{\begin{eqnarray}}
\def \eea{\end{eqnarray}}
\def \no {\nonumber}
\def \tr {{\rm Tr}}
\def \g {\gamma}
\def \tm {\mbb{T}}
\def \ha {\fr{ 1}{ 2}}
\def \half {\fr{ \trm{1}}{\trm{2}}}
\def \s {\sigma}
\def \vp {\varphi}
\def \td {\tilde}
\def \z {\zeta}
\def \H {{\rm H}}
\def \Tr {{\rm Tr}}
\def \ep {\epsilon}
\def \bp {\begin{pmatrix}}
 \def \emp {\end{pmatrix}}
\def \ef {\end{document}}
\def \del {\partial}
\def \G {\Gamma} \def \ha { { 1 \ov 2}}  \def \tg  {\td \Gamma} \def \m {\mu}
 \def \tdb {\bar }
\def \lm {Lam\'e\ }
 \def \tdb {\bar }
 \def\Z{\mathbb{Z}}
 \def\K{\mathbb{K}}
\def \ed {\end{document}}

 \newcommand{\ads}{AdS_5\times S^5}
 \def \sn {{\rm sn}} \def \bK {{\mathbb{K}}}

\let\5=\overline

\overfullrule=0pt
\parskip=2pt
\parindent=12pt
\headheight=0in \headsep=0in \topmargin=0in \oddsidemargin=0in

\vspace{ -3cm} \thispagestyle{empty} \vspace{-1cm}
 \vspace{-1cm}
\begin{flushright} 
\end{flushright}
\begin{center}
{\Large\bf Semiclassical short strings in $AdS_5 \times S^5$ }

 \vspace{0.8cm} {
  M.~Beccaria$^{a,}$\footnote{matteo.beccaria@le.infn.it},
  G.~Macorini$^{a,}$\footnote{guido.macorini@le.infn.it}, and
  A.~Tirziu$^{b,}$\footnote{atirziu@aps.org}}\\
 \vskip  0.5cm

\small
{\em
$^{a}$
Physics Department, Salento University and INFN, 73100 Lecce, Italy

 \vskip 0.05cm
$^{b}$ American Physical Society, 1 Research Road,
Ridge, NY 11961, USA
 }

\normalsize
\end{center}

 \vskip 0.8cm

 \begin{abstract}
 We present results for the one-loop correction to the energy of
 a class of string solutions in $AdS_5\times S^5$ in the short string limit.
 The computation is based on the observation that, as for rigid spinning string elliptic
solutions, the fluctuation operators  can be put into the single-gap  Lam\'e form.
Our computation reveals a remarkable universality of the
 form  of the energy of  short semiclassical strings. This may help to
 understand better the structure of the strong coupling expansion
  of the anomalous dimensions of dual gauge  theory operators.
\end{abstract}

\newpage

\tableofcontents

\renewcommand{\theequation}{1.\arabic{equation}}
 \setcounter{equation}{0}
\setcounter{footnote}{0}
\setcounter{section}{0}

\def \N {{\cal N}}\def \E {{\cal E}}
\def \PP  {{\mathscr{P}}}

\section{Introduction}

The energy of states of type IIB superstring propagating in \adss are dual to planar $\cal N$=4 SYM anomalous dimensions, and are interesting observable quantities depending on the string
tension  and various conserved charges. In principle, they are captured by the
Thermodynamic Bethe Ansatz \ci{reviews} which is well under control due to the integrability of the theory.
Nevertheless, it is important to identify specific limits where explicit analytical results can be given.

One such limit is the semiclassical expansion which is worked out in the limit when the conserved charges
scale as the string tension (or 't Hooft coupling) $\frac{\sql}{2 \pi} \gg 1$.
It is expected that in this limit  the TBA
prediction matches exactly the 1-loop string correction to the energies \cite{grom}.
One expects
fluctuation frequencies from algebraic curve  approach to match  the
ones found directly from string quadratic Lagrangian (both are
perturbations of solutions of same equations), and then one can argue
\cite{gv} that the string result interpreted as sum of fluctuation frequencies
extracted from algebraic curve exactly matches the strong coupling
expansion of TBA equations in same semiclassical limit.
The details of the matching  between the  algebraic curve description of the  semiclassical solutions
and their string $\sigma$-model presentation  remain still to be fully
  clarified.

Besides, one can extract from the semiclassical calculation a special short string limit
where the string configuration probes a small sized region of \adss. This is particularly interesting
in order to recover the $PSU(2,2|4)$ multiplet structure at the string level, including quantum corrections.

Technically,
apart from
 rational rigid   string solutions with
fluctuation Lagrangian containing
constant coefficients \ci{ft03},  the  quantum field theory
 computation  of one-loop string energies is complicated
 by mixed-mode
 fluctuation operators which are
  second order matrix two-dimensional
differential  operators with explicitly coordinate-dependent  coefficients.
In the case of a string which is folded in $AdS_{5}$ and rotates around its center,
the one-loop energy correction is related to the functional determinant of these operators. They can be computed
exactly since a reduction to Lam\'e integrable one-dimensional spectral problems is possible \cite{bd}.
A major simplification of the folded string case is that, on physical grounds, one expect to find a reference
frame where the fluctuation problem is static. A more difficult case is that of so-called pulsating string configurations.
In such a case one has, for instance, a string stretched along a parallel of $S^{5}$ which sweeps the sphere changing
its latitude bouncing back and forth aroung one of the poles. The fluctuation problem becomes intrinsically
time-dependent and a different formalism is required for the computation of the one-loop energy.

In this paper, we shall present the main steps of such a computation following the general semiclassical method
  of quantization of time-periodic solitons   \ci{dhn,viced}.

\renewcommand{\theequation}{2.\arabic{equation}}
 \setcounter{equation}{0}
\setcounter{section}{1}

\section{Pulsating string  in $\mathbb{R}\times S^{2}$}

\begin{figure}[hbtp]
\begin{center}
\includegraphics[scale=0.6]{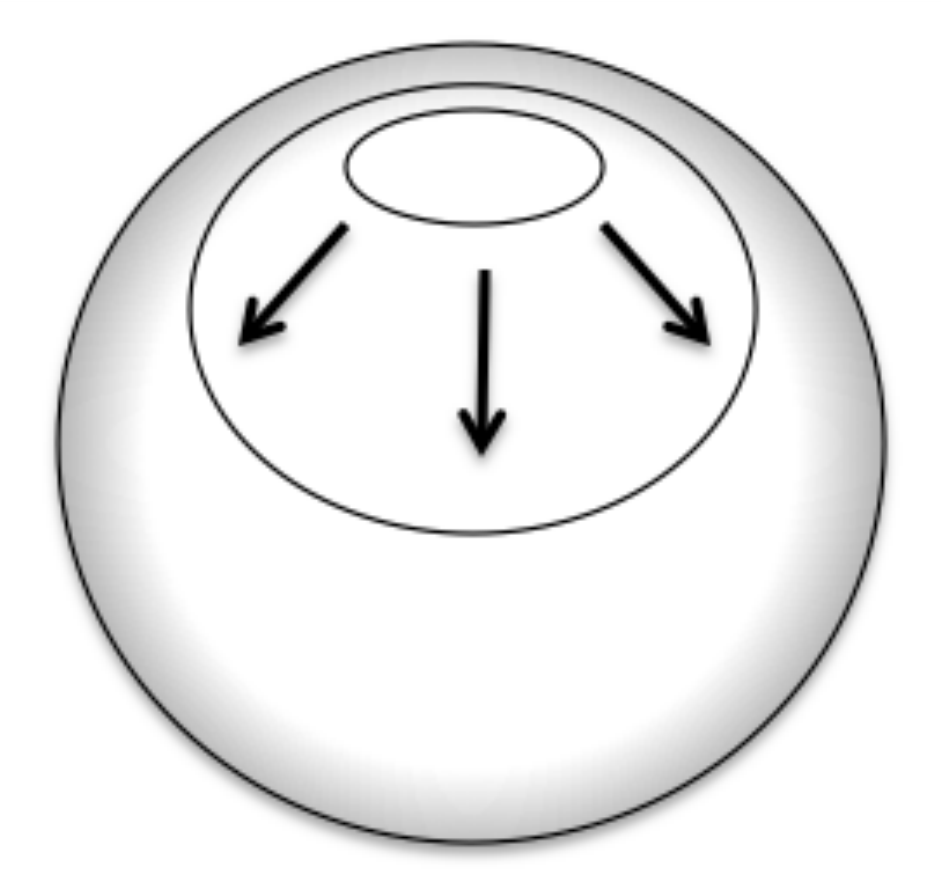}
 \caption{A qualitative picture of the pulsating string motion.}
 \end{center}
\end{figure}

We want to treat the classical   string solution
representing a pulsating string  in $\mathbb{R}\times S^{2}$. The motion of the string is depicted in
Figure~(1)~\ci{gkp,m03,krut}. We work in conformal gauge and start from the Ansatz for the
bosonic string degrees of freedom in $\mathbb{R} \times S^2$  ($m=1,2,...$)
\be
t = \kappa\,\tau, \qquad \psi=\psi(\tau), \qquad \phi=m\,\sigma\ , \qquad
ds^{2} = -dt^{2}+d\psi^{2}+\sin^{2}\psi\,d\phi^{2}  \ .\la{yllq}
\ee
The equation of motion and the  conformal gauge constraint (which
 implies the former for $\dot\psi\not=0$)
are
\be \ddot\psi+m^{2}\,\sin\psi\cos\psi = 0 \ , \ \ \ \ \ \
\dot\psi^{2}+m^{2}\,\sin^{2}\psi = \kappa^{2} \ .\la{llq}
\ee
The solution with $\psi(0)=0$ can be written in terms of the Jacobi elliptic function
(to  have a time-periodic solution we need to assume  $\kappa < m$, compatible with the later short string limit
) \cite{ww}
\be\la{soll}
\sin\psi(\tau) = \frac{\kappa}{m}\,{\rm sn}\Big(m\,\tau\,|\,\frac{\kappa^{2}}{m^{2}}\Big),
\qquad |\sin\psi| \le \sin\psi_{0} = \frac{\kappa}{m}.
\ee

The energy and the oscillation  number $N= \frac{\sqrt{\lambda}}{2\pi}\oint d\psi \ \dot \psi $
(which is the adiabatic invariant associated to $\psi$) are
\ba
\mathcal E_0 &=& \frac{E}{\sqrt\lambda} = \kappa\ ,\la{pqo} \\
\mathcal N &=& \frac{N}{\sqrt\lambda} = \int^{2\pi}_0 \frac{d\psi}{2\pi} \sqrt{\kappa^{2}-m^{2}\,\sin^{2}\psi} =
\frac{2m}{\pi}\Big[\Big(\frac{\kappa^{2}}{m^{2}}-1\Big)\,\mathbb{K}\Big(\frac{\kappa^{2}}{m^{2}}\Big)
+\mathbb{E}\Big(\frac{\kappa^{2}}{m^{2}}\Big)\Big]\ ,
\la{wpqo}\ea
where $\frac{\sql}{2
\pi}$ is string tension, and
 ${\mathbb K}$ and ${\mathbb E}$ are the usual elliptic functions \cite{ww,bd}. The expansion of $\mathcal N$  for  small $\kappa$ gives
\be
\mathcal N = \frac{\kappa ^2}{2 m}+\frac{\kappa ^4}{16 m^3}+\frac{3 \kappa ^6}{128 m^5}+\dots \ .
\la{nen}
\ee
Thus the  short string  (${\mathcal N} \to 0$)
 expansion of  the classical energy  reads
\be
\mathcal E_0(\mathcal N) = \sqrt{2\,m\,\mathcal N}\,\Big(
1-\frac{\mathcal N}{8 m}-\frac{5 \mathcal N^2}{128 m^2}+\dots
\Big). \la{ses}
\ee

\subsection{Quadratic fluctuations around the pulsating background}

In the semiclassical approach, one computes the one loop correction to the energy
starting from the quadratic fluctuations around the classical solution.
The bosonic  fluctuations in conformal gauge describe
 two mixed modes. They can be
decoupled  by exploiting the Virasoro constraints. An alternative equivalent approach
is based on the static gauge (see for instance \ci{bd}).

The conformal gauge fluctuations in $AdS_5$  describe
 a free massless  ghost field and  four free  massive fields
with  mass $ \kappa$.
(here $k=1,2,3,4$;\ \  $\del_a \del^a = - \del_\tau^2 + \del_\s^2$)
The Lagrangian for the  five  $S^5$ fluctuations ($ \xi,\eta,  z_1,z_2,z_3 $)
 is
\ba
&& L_{S}^{(2)} =
  \frac{1}{2}(\dot \xi^2-\xi'^2-M_\xi^2\,\xi^2) + \frac{1}{2}(\dot \eta^2-\eta'^2-M_\eta^2\,\eta^2)
 +\ m\,\cos\psi\,( \xi\,\eta'-\xi'\,\eta) \cr
 && \ \ \ \ \ \ \ \ \ \ \ \ \ \ +\
\frac{1}{2}(  \dot z_i^2  -   z'_i{}^2     - M^2 z_i^2    ) \ , \la{laga}
\ea
with the following background-dependent masses
\ba
M^2    =\kappa^{2}-2m^{2}\,\sin^{2}\psi, \qquad
M_\xi ^2    = \kappa^{2}+m^{2}\cos(2\psi) \ , \qquad M_\eta^2 =m^{2}\,\cos(2\psi)   \ .  \la{mas}
\ea
The  coupled system  $(\xi,\eta$) can be shown to be
equivalent  to a decoupled system of one  massless mode and
of the  massive mode with the Lagrangian
\ba\la{ggg}
L=
\frac{1}{2}(\dot g^2-g'^2-\td M^2 \, g^2) \  , \ \ \ \ \ \ \ \ \ \ \ \
\td M ^2= \kappa^{2}\big(1-\frac{2}{\sin^{2}\psi}\big)   \ . \ea
Yet another equivalent fluctuation action follows  also  from  the Pohlmeyer
reduction approach \cite{Iwashita:2010tg}.


The general fermionic fluctuation Lagrangian can be found, e.g., in \ci{ft02,bd}. Fixing $\kappa$ symmetry
as usual with  $\theta^1 = \theta^2$, and
after some standard manipulation, we can write the (squared) fermionic fluctuation operators
for both chiralities as
 \be\la{sqf}
\tilde D_F^{2}{}_\pm  = \partial_{\tau}^{2}-\partial_{\sigma}^{2}+ M^2_\pm \ , \ \ \ \ \ \ \ \ \ \ \ \
M^2_\pm= \dot\psi^{2}\pm i\,\ddot\psi.
\ee
Ultraviolet finiteness is easily checked by computing the supertrace of the squared mass matrix. The signed
sum of squared masses turns out to be proportional to  $\sqrt{-g}\,R^{(2)}$  , {\em i.e.}, the  Euler density of  the induced metric as
expected  on general grounds \ci{dgt}.
After integration over the 2-space, one finds a vanishing result for the
  cylinder topology which is appropriate for the string configuration under study.

\subsection{Fluctuation operators in Lam\'e  form}

We now show that all the obtained one-dimensional spectral problems can be put in
standard 1-gap Lam\'e form (see for instance the detailed discussion in \cite{bd}). This is important and means that all their relevant properties can be
worked out exactly.
Since the fluctuation potentials are independent of the spatial coordinate $\sigma$ for the
pulsating string solutions, we Fourier decompose all fields according to
$X(\tau, \sigma) = X(\tau)\ e^{in\sigma}$, so that $- \del_\tau^2 +
 \del_\s^2  +  M^2(\tau)\to  - \del_\tau^2 +    M^2(\tau) - n^2 $.
Depending on the form of the mass term (i.e. potential) $M^2(\tau)$,
we  find three types of Lam\' e operators, which we  discuss
below.

\subsubsection{Type I operator}

The operator associated to the three $S^{5}$ modes $z_i$ in \rf{laga}
 which have  mass $M^{2} = \kappa^{2}-2m^{2}\sin^{2}\psi$
is
\be\la{qqpa}
{\cal O}_{I} = -\partial_{\tau}^{2}+2m^{2}\sin^{2}\psi-\kappa^{2}-n^{2}\ .
\ee
For the pulsating string, Eq.~\rf{soll}, it  can be written as
\ba
&&{\cal O}_{I} = m^{2}\,\Big[-\partial^2_x+\  2k^2\,{\rm sn}^2(x\,|\,k^2)-\Lambda\Big], \la{o1} \\
&&
x = m\,\tau, \qquad {k^2 = \frac{\kappa^2}{m^2}}, \qquad \Lambda =
 \frac{\kappa^{2}+n^{2}}{m^{2}}  \  ,  \la{xxx} \nonumber
\ea
which is of  the single-gap Lam\'e form.

 \subsubsection{Type II operator}

{Next, consider the $S^{5}$ mode in \rf{ggg}
 with mass  $\td M^{2} = \kappa^{2}\big(1-\frac{2}{\sin^{2}\psi}\big)$, i.e.
with the associated operator
 \be \la{oqpq}
 {\cal O}_{II} = -\partial_{\tau}^{2}+\frac{2\kappa^{2}}{\sin^2\psi}-\kappa^{2}-n^{2} =
 m^{2}\,\Big[-\partial^2_x+\  2\,{\rm ns}^2(x\,|\,k^2)-\Lambda\Big]  \ ,
 \ee
 where we have used the definitions \rf{o1}.
Taking into account  the identity,
$
{\rm ns}(z\,|\,k^{2}) = k\,{\rm sn}(z+i\,\mathbb{K}'\,|\,k^{2})
$,
we have (we use the standard notation
${\mathbb K}^\prime(k^2)\equiv {\mathbb K}(1-k^2)$)
\ba\la{topa}
{\cal O}_{II} &=& m^{2}\,\Big[-\partial^2_{x}+\  2\,k^{2}\,{\rm sn}^2(x\,|\,k^2)-\Lambda\Big],\\
x &\equiv&  m\,\tau+i\,\mathbb{K}^\prime \quad, \quad k=\frac{\kappa}{m}\quad, \quad \Lambda = \frac{\kappa^{2}+n^{2}}{m^{2}}\quad , \nonumber
\ea
which is again of the single-gap Lam\'e form.}

\subsubsection{Type III operator}


The fermion fluctuation operator in  \rf{sqf}
 with the  mass {$M^2_\pm = \dot\psi^{2}\pm i\, \ddot\psi$}
leads to
 \be
 {\cal O}_{III}^\pm = -\partial_{\tau}^{2}-\dot\psi^{2}\mp i\,\ddot \psi-n^{2}.
 \ee
After some manipulation of the elliptic functions, we can show that it can be written\ba
&& {\cal O}_{III}^\pm =
\bar{m}_\pm^{2}\,\Big[-\partial^2_x+2\,\bar{k}_\pm^{2}\,{\rm sn}^2(\bar{x}\,|\,\bar{k}_\pm^2)
-\Lambda\Big] \ ,  \la{kol}
\\
&& \bar x \equiv \bar{m}_\pm\,\tau+\frac{1}{2} {{\mathbb K}(\bar{k}^2_\pm)} \ ,
\qquad \bar m_\pm =\frac{m}{2}\Big(\sqrt{1-\frac{\kappa^2}{m^2}}\pm i\frac{\kappa}{m}\Big)\  , \la{pyy}\\
&&
\bar{k}_\pm^2=\pm 4\frac{ \frac{i\,\kappa}{m} \sqrt{1-\frac{\kappa^2}{m^2}}}
{\Big( {\sqrt{1-\frac{\kappa^2}{m^2}}} \pm {\frac{i\kappa}{m}}\Big)^2}\ , \qquad
 \Lambda = \frac{n^{2}}{\bar{m}_\pm^{2}+\bar{k}_\pm^2}\ .
\la{jjj2}
\ea
Thus
 we  again  find a fluctuation operator
  of the single-gap Lam\'e form.

\renewcommand{\theequation}{3.\arabic{equation}}
 \setcounter{equation}{0}
\setcounter{section}{2}

\section{Semiclassical quantization of time-periodic solutions of
integrable systems}

Let us consider a classical Hamiltonian system on a space  $X$ ($\dim X = 2n$)
 with a  Hamiltonian
$
H: T^{*}X\to \mathbb R.
$
Its quantum version will be  a self-adjoint operator $\widehat H$ on
 such that in the classical limit $\hbar\to 0$ it reduces to
$H$.
{\em Classical} integrability requires the existence of $n$
functions $F_{1}, \dots, F_{n}\in C(T^{*}X)$ such that:
(i) $ dF_{1}\wedge\cdots \wedge dF_{n}\neq 0, \ {\rm almost\  everywhere},$
(ii) $\{F_{i}, F_{j}\}=0,$  and  (iii)
$H = H(F_{1}, \dots, F_{n})$.
This implies that the level sets
define $n$-tori (Liouville tori) foliating $T^{*}X$ and
invariant under the Hamiltonian flow. This allows one  to define the action
variables $I_{i}$ parametrizing the
foil base and the angle variables $\varphi_{i}$, the  coordinates of the torus.
The weaker condition of {\it semiclassical} integrability requires the existence of quantum extensions
$\widehat F_{i}$ of $F_i$  ($\widehat F_{i}\stackrel{\hbar\to 0} {\lar} F_{i}$)
such that in addition to  the condition (i)   above  they satisfy
(ii') $ [\widehat F_{i}, \widehat F_{j}]=\mathcal O(\hbar^{3})$   and (iii') $
\widehat H = H(\widehat F_{1}, \dots, \widehat F_{n})+\mathcal O(\hbar^{2})$.
Note  that $\widehat H$  is well defined without ordering problems
because of the condition (ii').

The joint semiclassical  diagonalization   problem
\be
\widehat F_{i}\,\psi = f_{i}\,\psi+\mathcal O(\hbar^{2})\ ,\la{jo}
\ee
can be solved by a WKB-like approximation which require
the following  Bohr-Sommerfeld-Maslov (BSM) quantization condition  \cite{Voros}
\be
\frac{1}{2\pi\hbar}\int_{\gamma_{i}} {p}\cdot d  q = N_{i}+\frac{\mu_{i}}{4}+\mathcal O(\hbar), \qquad
i=1, \dots, n \ ,
\ee
where the  integers $N_i$  thus define  the action variables,
 $\{\gamma_{i}\}$ is  a basis of  cycles of a Liouville torus,
and the Maslov indices $\mu_{i}$ take into
account the critical points of the cycles.

If the classical invariant torus has only $p<n$ non trivial cycles, then the  BSM
quantization condition must be modified in order to
take into account the fluctuations transverse to the codimension $p$
invariant torus. It becomes
\be
\frac{1}{2\pi\hbar}\int_{\gamma_{k}} {p}\cdot d q =N_{k}+\frac{\mu_{k}}{4}
+\sum_{\alpha=p+1}^{n}\big(n_{\alpha}+\frac{1}{2}\big)\,\frac{\nu_{\alpha}^{(k)}}{2\pi}
+\mathcal O(\hbar),\quad \begin{array}{l} \ \ \ \ \ k=1, \dots, p, \\ \ \ \ \ \ n_{\alpha}\ll N_{k}\end{array}
\la{hhh}\ee
The {\it stability angles}
$\nu_{\alpha}^{(k)}$ can be found  by studying
 the stability of small fluctuations
around the invariant torus (the condition $n_{\alpha}\ll N_{k}$ is necessary
in order  to  be able to use
 the linearised analysis).

In the case of semiclassical quantization of
finite $g$-gap solutions of string theory,
 one starts with a   classical energy as a function of the  action  variables
 and then simply shifts them   according to the  BSM
 quantization
conditions \cite{viced}
\ba
 &&E = E_{cl}\Big(N_{1}\hbar+\frac{\hbar}{2}+\hbar\sum_{\alpha=g+2}^{\infty}\big(n_{\alpha}+\frac{1}{2}
\big)\frac{\nu_{\alpha}^{(1)}}{2\pi}, \dots,
\nn
&&\qquad\qquad
N_{g+1}\hbar+\frac{\hbar}{2}+\hbar\sum_{\alpha=g+2}^{\infty}\big(n_{\alpha}+\frac{1}{2}
\big)\frac{\nu_{\alpha}^{(g+1)}}{2\pi}\Big)+\mathcal O(\hbar^{2})\ . \la{eqe}
\ea
In particular, for the ground state  ($n_\alpha=0$)
 of a  1-gap superstring time-dependent
solution
of period $\T$, we can write (here $\hbar=\frac{1}{\sqrt\lambda}$, $\mathcal N = \frac{N}{\sqrt\lambda},
\ \mathcal E = \frac{E}{\sqrt\lambda}$)
\be
\mathcal E = \mathcal E_{cl}(\mathcal N) + \frac{1}{2\sqrt\lambda}\frac{1}{\T}\sum_{\nu_{s}>0 }\nu_{s}+
\mathcal O(\frac{1}{(\sqrt\lambda)^{2}}) \ . \la{pou}
\ee
Here $\T$ is the period of the solution which is the inverse of $\frac{d E}{d N}$.


In general, in an integrable system,
 the stability angles  may  be computed starting from the problem of evolution of a small
perturbation which is controlled by the non-linear superposition principle associated with
B\"acklund
 transformations. The same construction can be interpreted as the addition of an infinitesimal
  cut
to a  finite cut solution of the corresponding integral equations
implied by the Bethe equations. Also, a third point of view is that of considering a genus $g+1$ algebraic
curve infinitesimally near its  genus $g$  degeneration point.
In the case of the pulsating string, this sophisticated discussion will simply boil down to the
nice spectral properties of the Lam\'e equation. 


\renewcommand{\theequation}{4.\arabic{equation}}
 \setcounter{equation}{0}
\setcounter{section}{3}

\section{One-loop correction to energy}

In general, given the 1-d spectral problem with a periodic potential
\begin{equation}\label{rder}
 \big[-\del_x^2 +V(x) \big]\,f(x) =\Lambda\, f(x) \ , \ \ \ \ \ \ \ \ \ \ \
 V(x+\T)=V(x)  \ ,
\end{equation}
its two independent solutions  $f_\pm (x)=e^{\pm i \, p(\Lambda) \, x}\,
\chi_\pm (x), \ \  \chi_\pm (x+\T)= \chi_\pm (x )$   satisfy
 \begin{eqnarray}
 f_\pm(x+\T)=e^{\pm i\nu }\ f_\pm(x) \ , \ \ \ \ \ \  \nu = p \T  \ ,
 \label{ntum}
 \end{eqnarray}
where $\nu$ is the ``stability angle'' and $p$ is the ``quasi-momentum''
  (in general,  $p$  is a function of $\Lambda, \T$ and a functional of $V$).

For the pulsating string in $\mathbb{R}\times S^{2}$
 the period is $\T=\frac{4\K}{m}$.
The short string limit  is the small $\kappa$ limit, in which the semiclassical oscillation
parameter $\mathcal{N}$  is small.
Below we shall
consider the positive of the  two possible  stability angles differing by sign
 (see \cite{hass}).
 Since in the present case the $AdS$ time $t$ and 2d  time $\tau$ are related as in
\rf{yllq},
i.e. $t = \k \tau$, there will be similar proportionality of  the periods,
and   the space-time energy  and the 2d energy will be related by
\be \la{repa}
E_{\rm spacetime} = \frac{1}{\k} E_{2d}  \ . \ee

\subsection{Stability angles}

The $4$  massless $ AdS_5$ fluctuations   have  simply the stability angle
\be
\nu_{_{AdS_{5}}}  = 4\K\,\sqrt{k^{2}+\frac{n^{2}}{m^{2}}} \ , \ \ \ \ \ \ \ \
k\equiv \frac{ \kappa}{m}   \ . \la{io}
\ee
As we have shown, the $S^{5}$ bosonic fluctuations (both Type I \rf{qqpa} and Type  II \rf{oqpq}) are associated with the
 standard Lam\'e equation, and therefore the stability angle is
\be\la{cha}
&&\nu_{_{S^{5}}} = \pm 4\,\K\,\left(i\,\Z(\alpha\,|\,k^{2})+\frac{\pi}{2\K}\right) \equiv
 \pm 4\,\K\,i\,\Z(\alpha\,|\,k^{2})\ ,
\\
&&{\rm sn}(\alpha\,|\,k^{2}) = \sqrt\frac{1+k^{2}-\Lambda}{k^{2}} =
\frac{1}{k}\sqrt{1-\frac{n^{2}}{m^{2}}}\ .
\ee
We shall fix the sign in \rf{cha}   by the condition $\nu>0$. Finally, in the case of the
 fermionic fluctuation operator  the  expression for
  the stability angle is
\begin{equation}
\nu_{_F} = \pm 4\,i\,\mathbb{K}\,\Big[
\frac{1}{2}\,Z(\alpha({\beta})\,|\,k^{2})+
i\,\sqrt{\beta}\,\sqrt{1+\frac{16\,\beta\,k^{2}}
{(1-4\beta)^{2}}}\
\Big],
\end{equation}
where
\begin{equation}
\alpha(\beta) = {\rm cn}^{-1}\Big(-\frac{1+4\beta}{1-4\beta}\,|\,k^{2}\Big)\ , \ \ \ \ \ \ \ \
\   \beta=\frac{n^{2}}{m^{2}}  \ .
\end{equation}

Let us now combine the above  stability angles expanded in powers of
 $\k= { k  m}$   with proper multiplicities and
signs as they should
appear in the 1-loop correction to the energy in \rf{pou}
\ba
&&  {\nu_{n} = 4\times (\nu_{_{AdS_{5}}}+\nu_{_{S^{5}}})-8\times\nu_{_F} } \no \\
&&
\ \ \ =\frac{4 \pi \k^2 m }{n\,(m^2 -4 n^2)}-\frac{\pi  \k^4
 (2 m^8-28 m^6 n^2+133 m^4 n^4-128 m^2 n^6+48 n^8)}{2m
   n^3 (m^2-4 n^2)^3 (m^2-n^2)} \nonumber \\
   &&\ \ \  +\ \frac{\pi  \k^6  }{16 m^3(m^2-n^2)^2 (m^2
   n-4 n^3)^5}\Big(8 m^{16}-180 m^{14} n^2+1705 m^{12} n^4-8772
   m^{10} n^6+25883 m^8 n^8\nonumber\\
   && \ \ \ \ \ \ \ \ \ \ \ \ \-\ 35456 m^6 n^{10} +25824 m^4 n^{12}-13824 m^2 n^{14}+3840 n^{16}\Big)
   +...\ . \la{summa}
\ea
As a check,
 we observe that the sum over  $n$  of this combination is
  convergent at large $n$. Setting $m=1$ and summing up all the contributions   we get
from \rf{pou}    the following expression for the
1-loop correction to the string energy  (taking into account the relation  \rf{repa} valid
in the static gauge $t=\k \tau$)
\ba
\mathcal E_1=
\frac{1}{2 \T \k } \sum_{n=-\infty}^{\infty} {\nu_{n}} &=& 2 +\kappa  (1-4 \log 2)+ \frac{1}{8}
\kappa ^3
\Big({3 \zeta_3}+ {1}+4 \log 2 \Big) \nonumber \\
&&+ \frac{ 1}{4} \kappa ^5 \Big(-\frac{63 \zeta_3
   }{16}-\frac{15 \zeta_5}{16}+\frac{7}{32}+ {\log 2}\Big)+O(\kappa
    ^7)   \ . \la{ghjk}
\ea

In general, we can organize the short string expansion of the energy as
\be
&&E = E\Big(\frac{N}{\sqrt\lambda}, \sqrt\lambda\Big) = \sqrt\lambda\,\mathcal E_{0}(\mathcal N)+
\mathcal E_{1}(\mathcal N)+\frac{1}{\sqrt\lambda}\,\mathcal E_{2}(\mathcal N)+... \  ,
\la{kouu}\\
&&
\mathcal E_{k}= \sqrt{2\mathcal N}\ \Big(a_{0k}+a_{1k}\,\mathcal N+a_{2k}\,\mathcal
N^{2}+... \Big)
 + c_{0k}+c_{1k}\,\mathcal N+.... \ . \la{jrkp}
\ea
where $c_{nk}$ are coefficients of  ``non-analytic''
 terms \cite{rt}.
Using \rf{nen},\rf{ses}  and \rf{ghjk} we find that
for the pulsating string in  $\mathbb{R}\times S^{2}$
\ba
\mathcal E_{0} &=& \sqrt{2\mathcal N}\Big(1-\frac{1}{8}\mathcal N-\frac{5}{128}\mathcal N^{2}
+... \Big), \la{kla}\\
E_1\equiv \mathcal E_{1} &=& 2+\sqrt{2\mathcal N}\Big[
 1-4 \log 2+\Big(\frac{3}{2}\log 2+\frac{3}{4} \zeta_{3}+\frac{1}{8}\Big)\mathcal N
  \no \\
 &&  + \  \Big(
\frac{25}{32}\log 2-\frac{135}{32}\zeta_{3}-\frac{15}{16}\zeta_{5}+\frac{11}{128}
\Big)\mathcal N^{2}+...\Big]. \la{kqq}
\ea
Therefore, the energy can be re-written
in terms of $N$ and the string tension as follows
\ba
&&E = \sqrt{2N\sqrt\lambda}\ \Big(a_{00}+\frac{a_{10}N+a_{01}}{\sqrt\lambda}+
...\Big)  + c_{01}+ ... \ , \la{ert} \\
&&a_{00}=1, \qquad a_{10}=-\frac{1}{8}, \qquad a_{01} = 1-4\log 2, \qquad c_{01}=2\ ,\ \ \  ...
\la{yyyi}
\ea

\renewcommand{\theequation}{5.\arabic{equation}}
 \setcounter{equation}{0}
\setcounter{section}{4}

\section{Generalizations and concluding remarks}

In this paper we extended  the investigation \cite{bd}
of the exact structure of  one-loop correction to energy of an  important class of classical string solutions
in \adss  expressed in terms of
elliptic functions.
The elliptic  solution considered  in  \cite{bd} was the folded spinning  string in $AdS_5$
for which it was shown that the  quadratic fluctuation operators can be put into  the standard single-gap L\' ame form.
This is an important feature allowing to compute  the one-loop correction to the string energy
exactly for any value of semiclassical spin parameter $\S$.
Here we have extended the calculation to the pulsating string in $\mathbb R\times S^{2}$. Additional
details as well as the extensions to the pulsating string in $AdS_3$ and  the folded spinning string in
 $\mathbb R\times S^{2}$ can be found in \cite{Beccaria:2010zn}.

 In all cases where
there is only one charge/adiabatic invariant besides the energy,  namely, an  oscillator
 number or spin (in $S^5$ or $AdS_5$),
the  fluctuation operators  can be decoupled   and
 put into  a single-gap Lam\'e type form.

We focused on the expansion of the one-loop  energies
in the  limit of small values of the semiclassical parameters  corresponding to small size of the string.
In this limit  the string  probes
only small region of  \adss so  its energy should start with   the  standard
 flat-space form plus corrections due to curvature.
This limit
 may provide further insight into the structure of  strong-coupling corrections to dimensions of ``short''  dual  gauge
theory operators for which the  ``wrapping'' contributions are important \ci{tt,rt}.

The semiclassical approximation is based on assumption  that $\sql \gg 1$ with
semiclassical parameters characterizing the various solutions (see \cite{Beccaria:2010zn}, for the precise definitions)
like $ \S= \frac{ S }{ \sql}, \ \J = \frac{ J }{\sql}$ or $\N= \frac { N }{ \sql}$
fixed, so that  $S, J$ or $N$ are   formally  large. An intriguing conjecture is that, taking  the  ``short-string''  limit  in which $\S, \J, \N \to 0$,   one may assume that the limit ``commutes''   with the large $\sql$ limit. If so, the semiclassical
analysis  may shed  light on the form of
the quantum string energies with  fixed values of  the spins and oscillation numbers $(S,J, N)$.
While this is only a conjecture, the study of  the ``short-string'' limit
provides some  qualitative  information on the structure of the large tension expansion  of quantum   string energies or
strong-coupling expansion of dimensions of dual
gauge-theory operators.

We  summarize below the results for the ``short-string'' (small spin or oscillation number) expansion of the classical
$E_0$  and one-loop $E_1$
energies of the four basic  elliptic \adss solutions analysed in  \cite{bd} and here:
folded  spinning strings  in   $\mathbb R\times S^{2}$   and  $AdS_3$,
and pulsating circular strings in  $\mathbb R\times S^{2}$   and  $AdS_3$.
We recall our notation:
$E=E_0 + E_1 + ...$, $E_0 = \sql  \mathcal E_{0} $,  $E_1= \mathcal E_{1}$.
Also,  the non-zero  spin of the folded string in $S^2$ is   $J_2\equiv J$.


\medskip\noindent
\underline{\em Folded spinning string in $\mathbb R\times S^{2}$  }
\ba
\mathcal E_{0} &=& \sqrt{2\,\mathcal J}\,\Big(1+\frac{1}{8}\,\mathcal J
+\frac{3}{128}\mathcal J^{2}+\dots\Big), \no \\
   E_{1} &=& 2+\sqrt{2\,\mathcal J}\,\Big[2-4\,\log 2 + \Big(
-\frac{1}{2}-\frac{3}{2}\log 2+\frac{3}{4}\zeta_{3}
\Big)\,\mathcal J  \no  \\
 & &  \ \ \ \ \  +    \Big(\frac{1}{64}-\frac{15}{32}\log 2+\frac{51}{32}\zeta_{3}-\frac{15}{16}\zeta_{5}\Big)\,\mathcal J^{2}
+\dots
\Big], \nonumber \\
E &=& \sqrt{2J  \sqrt\lambda}\Big(1+\frac{\frac{1}{8}J+2-4\log 2}{\sqrt\lambda}+\dots\Big)+2+\dots
\la{ro1}
\ea

\medskip\noindent
\underline{\em Folded  spinning  string in $AdS_{3}$}
\ba
\mathcal E_{0} &=& \sqrt{2\,\mathcal S}\,\Big(1+\frac{3}{8}\,\mathcal S-\frac{21}{128}\mathcal S^{2}
+\dots\Big),\no  \\
   E_{1} &=& 1+\sqrt{2\,\mathcal S}\,\Big[\frac{3}{2}-4\,\log 2 + \Big(
-\frac{23}{16}+\frac{3}{2}\log 2+\frac{3}{4}\zeta_{3}
\Big)\,\mathcal S    \no \\
 & & \ \ \ \ \ \   +
\Big(
\frac{689}{256}-\frac{63}{32}\log 2-\frac{15}{32}\zeta_{3}-\frac{15}{16}\zeta_{5}
\Big)\,\mathcal S^{2}+\dots
\Big], \nonumber \\
E &=& \sqrt{2S\sqrt\lambda}\Big(1+\frac{\frac{3}{8}S+\frac{3}{2}-4\log 2}{\sqrt\lambda}+\dots\Big)+1+\dots
\la{ro2}
\ea

\medskip\noindent
\underline{\em Pulsating string in $\mathbb R\times S^{2}$}
\ba
\mathcal E_{0} &=& \sqrt{2\,\mathcal N}\,\Big(1-\frac{1}{8}\,\mathcal N-\frac{5}{128}\,\mathcal N^{2}+\dots\Big),\no  \\
  E_{1} &=& 2+\sqrt{2\,\mathcal N}\,\Big[1-4\,\log 2 + \Big(
\frac{1}{8}+\frac{3}{2}\log 2+\frac{3}{4}\zeta_{3}
\Big)\,\mathcal N    \no \\
 & & \ \ \ \ \    +   \Big(
\frac{11}{128} + \frac{25}{32}\log 2-\frac{135}{32}\zeta_{3}-\frac{15}{16}\zeta_{5}
\Big)\mathcal N^{2}+\dots
\Big],\nonumber \\
E &=& \sqrt{2N\sqrt\lambda}\Big(1+\frac{-\frac{1}{8}N+1-4\log 2}{\sqrt\lambda}+\dots\Big)+2+\dots
\la{pu1}
\ea

\medskip\noindent
\underline{\em Pulsating string in $AdS_{3}$}
\ba
\mathcal E_{0} &=& \sqrt{2\,\mathcal N}\,\Big(1+\frac{5}{8}\,\mathcal N-\frac{77}{128}\,\mathcal N^{2}+\dots\Big),\no
 \\
  E_{1} &=& 1+\sqrt{2\,\mathcal N}\,\Big[
\frac{5}{2}-4\log 2+\Big(
-\frac{37}{8}+\frac{5}{2}\log 2+\frac{3}{4}\zeta_{3}
\Big)\,\mathcal N    \no \\
 & & \ \ \ \ \   +
\Big(
\frac{3915}{256}-\frac{231}{32}\log 2-\frac{117}{32}\zeta_{3}-\frac{15}{16}\zeta_{5}\Big)\,\mathcal N^{2}
+\dots
\Big],\nonumber \\
E &=& \sqrt{2N\sqrt\lambda}\Big(1+\frac{\frac{5}{8}N+\frac{5}{2}-4\log 2}{\sqrt\lambda}+\dots\Big)+1+\dots
\la{pu2}
\ea
We observe a remarkable universality of the  small charge expansion of the energy of
 all four elliptic solutions: the  leading terms with  transcendental  coefficients
($\log 2, \ \zeta_{3}, \ \zeta_{5}, ...$)  happen to have the same form.

Strings with lowest non-trivial  values of the charges  should correspond
to string states at the first excited level. Since these  should be dual  to members of the Konishi multiplet, they
should  have the same  anomalous dimension \ci{rt}.
The relationship between this intriguing universality, the  $PSU(2,2|4)$ structure of the Konishi multiplet, and the
validity of the above mentioned conjecture
remains to be clarified.


\section*{Acknowledgments }
We thank our collaborators A. A. Tseytlin and G. Dunne. We also thank V. Forini, N. Gromov,  M. Kruczenski,  R. Roiban  and
B. Vicedo for many useful discussions. 

 \bigskip


\end{document}